\documentclass[sigconf, table]{acmart}
\AtBeginDocument{%
  }

\usepackage{adjustbox}
\usepackage{cleveref}
\usepackage{tcolorbox}

\usepackage{listings}
\usepackage{xcolor}

\usepackage{hyperref}

\lstdefinelanguage{yaml}{
    keywords={true, false, null, yes, no},
    keywordstyle=\color{blue}\bfseries,
    basicstyle=\ttfamily\small,
    comment=[l]{\#},
    morecomment=[s]{/*}{*/},
    commentstyle=\color{gray}\itshape,
    stringstyle=\color{red},
    moredelim=[l][\color{black}]{---},
}

\renewcommand\footnotetextcopyrightpermission[1]{}
\acmConference[arXiv preprint]{Make sure to enter the correct
  conference title from your rights confirmation email}{preliminary work}{v1: March 2025}
\acmISBN{978-1-4503-XXXX-X/2018/06}

\begin{document}

\title[ContextGNN goes to \textsc{Elliot}: Towards Benchmarking Relational Deep Learning for Personalized Item Recommendation]{ContextGNN goes to \textsc{Elliot}: Towards Benchmarking Relational Deep Learning for Static Link Prediction (aka Personalized Item Recommendation)}


\author{Alejandro Ariza-Casabona}
\authornote{All authors contributed equally, and their names are reported in alphabetical order.}
\email{alejandro.ariza14@ub.edu}
\affiliation{%
  \institution{University of Barcelona, CLiC-UBICS}
  \city{Barcelona}
  \country{Spain}
}

\author{Nikos Kanakaris}
\authornotemark[1]
\email{kanakari@usc.edu}
\affiliation{%
  \institution{University of Southern California}
  \city{Los Angeles}
  \state{CA}
  \country{USA}
}

\author{Daniele Malitesta}
\authornotemark[1]
\email{daniele.malitesta@centralesupelec.fr}
\affiliation{%
  \institution{Université Paris-Saclay, CentraleSupélec, Inria}
  \city{Gif-sur-Yvette}
  \country{France}
}

\renewcommand{\shortauthors}{Ariza-Casabona et al.}

\begin{abstract}
Relational deep learning~\cite{DBLP:conf/icml/FeyHHLR0YYL24} (RDL) settles among the most exciting advances in machine learning for relational databases, leveraging the representational power of message passing graph neural networks (GNNs) to derive useful knowledge and run predicting tasks on tables connected through primary-to-foreign key links. The RDL paradigm has been successfully applied to recommendation lately, through its most recent representative deep learning architecture namely,  ContextGNN~\cite{DBLP:journals/corr/abs-2411-19513}. While acknowledging ContextGNN's improved performance on real-world recommendation datasets and tasks, preliminary tests for the more traditional static link prediction task (aka personalized item recommendation) on the popular Amazon Book dataset have demonstrated how ContextGNN has still room for improvement compared to other state-of-the-art GNN-based recommender systems. To this end, with this paper, we integrate ContextGNN within \textsc{Elliot}, a popular framework for reproducibility and benchmarking analyses, counting around 50 state-of-the-art recommendation models from the literature to date. On such basis, we run preliminary experiments on three standard recommendation datasets and against six state-of-the-art GNN-based recommender systems, confirming similar trends to those observed by the authors in their original paper. The code is publicly available on GitHub\footnote{\url{https://github.com/danielemalitesta/Rel-DeepLearning-RecSys}}.  
\end{abstract}



\maketitle

\section{Introduction and motivations}

As most of today's data sources, recommendation data is generally stored in relational databases with tables recording information about users, products and their multiple interconnections. Training machine learning models to perform prediction tasks on relational databases is challenging, as the existing approaches fail to capture the complex relations that exist among tables, thus requiring ad-hoc feature engineering procedures. 

To this end,~\citet{DBLP:conf/icml/FeyHHLR0YYL24} recently introduced a novel end-to-end deep representation learning paradigm, \textit{relational deep learning} (RDL), which re-interprets the table rows and the primary-to-foreign key links as the nodes and edges of a temporal and heterogeneous graph, respectively. Under this perspective, RDL can seamlessly exploit the representational power of message passing graph neural networks (GNNs) to derive useful knowledge patterns from relational databases and perform several prediction tasks. 

At the time of the current paper, the first concrete effort to leverage RDL for recommendation has been recently proposed through ContextGNN~\cite{DBLP:journals/corr/abs-2411-19513}, a novel recommendation system built on the RDL rationale, which addresses the limitations of current two-tower and pair-wise recommendation approaches. In the same paper, the authors tested ContextGNN on real-world tasks and datasets introduced in RelBench\footnote{\url{https://github.com/snap-stanford/relbench}}~\cite{DBLP:conf/nips/0001RHHHDFLYZHL24}, the main benchmarking framework for RDL. Then, the performance of ContextGNN was assessed on more traditional datasets and tasks, accounting for static link prediction and temporal next-item prediction. 

\textbf{Observation 1.} Although results on the latter demonstrate the vast predominance of ContextGNN to existing baselines and even though ContextGNN appears quite competitive against more traditional GNN-based recommender systems on static link prediction (also commonly known as personalized item recommendation), it is still not as powerful as more advanced and recent recommendation approaches. The authors of ContextGNN discussed possible future improvements to their framework, paving the way to more advanced techniques from the recent literature, proposing, among others, simplifications of the message passing procedure for recommendation~\cite{DBLP:conf/sigir/0001DWLZ020, DBLP:conf/cikm/MaoZXLWH21} and self-supervised learning strategies~\cite{DBLP:conf/sigir/WuWF0CLX21, DBLP:conf/iclr/Cai0XR23}.

\textbf{Observation 2.} Even acknowledging the urgency to build and test novel recommendation approaches capable of performing well on real-world datasets for concrete recommendation tasks~\cite{DBLP:journals/corr/abs-2411-19513}, we maintain the relevance of academic research focusing on more traditional tasks and datasets for personalized recommendation as an indispensable tool to foster advancements in recommendation systems designed for real-world applications. 

\textbf{Our contributions.} For this reason, with this paper, we decide to integrate ContextGNN\footnote{\url{https://github.com/kumo-ai/ContextGNN}} within \textsc{Elliot}~\cite{DBLP:conf/sigir/AnelliBFMMPDN21, DBLP:conf/recsys/AnelliMPBSN23}, a popular framework for rigorous and reproducible recommender systems evaluation. As in the original spirit of \textsc{Elliot}, the idea is to provide the end user with a ready-to-use tool that can extensively train and evaluate ContextGNN against a large number of state-of-the-art recommender systems from the literature (around 50) on any recommendation dataset through an easily-configurable YAML file.

In the remaining of the paper, we provide technical foundations on the ContextGNN model, designed for the task of personalized item recommendation. Then, we present our experimental study where we reproduce results from the ContextGNN paper~\cite{DBLP:journals/corr/abs-2411-19513} on three standard recommendation datasets (i.e., Gowalla~\cite{DBLP:conf/www/LiangCMB16}, Yelp 2018~\cite{DBLP:conf/sigir/0001DWLZ020}, and Amazon Book~\cite{DBLP:conf/www/HeM16}) and benchmark it against six state-of-the-art GNN-based recommendation systems (i.e., NGCF~\cite{DBLP:conf/sigir/Wang0WFC19}, DGCF~\cite{DBLP:conf/sigir/WangJZ0XC20}, LightGCN~\cite{DBLP:conf/sigir/0001DWLZ020}, SGL~\cite{DBLP:conf/sigir/WuWF0CLX21}, UltraGCN~\cite{DBLP:conf/cikm/MaoZXLWH21}, and GFCF~\cite{DBLP:conf/cikm/ShenWZSZLL21}), whose results were carefully reproduced in a recent paper~\cite{DBLP:conf/recsys/AnelliMPBSN23}. Finally, to facilitate future use of our tool, we provide guidelines on how to reproduce and benchmark ContextGNN within \textsc{Elliot} against other baselines and novel recommendation datasets. 

The aim of this paper is to release a tool that researchers and practitioners can easily use and contribute to reproduce and benchmark RDL recommendation systems on the traditional task of personalized item recommendation. The code to fully reproduce our experiments and run ContextGNN for more large-scale benchmark analyses is accessible on GitHub\footnote{\url{https://github.com/danielemalitesta/Rel-DeepLearning-RecSys}}.
\section{Background notions}

In this section, we present the necessary background notions regarding ContextGNN for static link prediction, namely, personalized item recommendation. We refer to~\cite{DBLP:journals/corr/abs-2411-19513} for a more in-depth mathematical formulation on the same topic.


\textbf{Problem definition.}  
The personalized item recommendation task aims to predict which items a user is most likely to interact with based on historical interactions. We formulate this problem as a link prediction task on a static user-item graph \( \mathcal{G} = (\mathcal{V}, \mathcal{E}, \varphi, \psi) \), where \( \mathcal{V} \) is the set of nodes, consisting of users \( \mathcal{U} \subset \mathcal{V} \) and items \( \mathcal{I} \subset \mathcal{V} \), while \( \mathcal{E} \subseteq \mathcal{U} \times \mathcal{I} \) represents the observed user-item interactions, such as purchases, clicks, or ratings, \( \varphi(v) \) denotes the node type, either user or item, and \( \psi(e) \) represents the interaction type. Given the observed user-item interactions in \( \mathcal{E} \), the goal is to predict a set of items \( \hat{\mathcal{I}}_u \subset \mathcal{I} \) that are most relevant to each user \( u \in \mathcal{U} \), thereby generating personalized recommendations. 

To better understand how recommendations are localized, the authors in~\cite{DBLP:journals/corr/abs-2411-19513} define the locality score \( s_k \), which quantifies the degree to which a user’s recommendations are based on their local subgraph (we adapt their formulation to static link prediction by dropping the temporal component in the graph):
\begin{equation}
    s_k =
    \frac{1}{|\mathcal{U}|}
    \sum_{u \in \mathcal{U}} \frac{|\mathcal{N}_k(u) \cap \mathcal{I} \cap \mathcal{I}_u|}{|\mathcal{I}_u|}
\end{equation}
where \( \mathcal{N}_k(u) \) is the \textit{k}-hop neighborhood of user \( u \), representing locally connected items and \( \mathcal{I}_u \) is the set of hidden true positive items for user \( u \), meaning items the user would likely interact with. On such a basis, ContextGNN employs a hybrid scoring mechanism to balance localized and global item recommendations through a pair-wise and two-tower representation score, respectively.

\textbf{Pair-wise representation score.}  
The pair-wise representation is applied when the item \( i \) lies within the user’s local subgraph \( \mathcal{N}_k(u) \). This approach enhances the model’s ability to capture localized preference patterns and repeat interactions. The computation follows these steps:

\begin{enumerate}
    \item \textsc{Subgraph construction}: A \textit{k}-hop subgraph \( \mathcal{G}_k(u) \) is extracted around the user \( u \), incorporating all directly and indirectly connected items up to \( k \) hops.
    \item \textsc{GNN-based representation learning}: A graph neural network (GNN) processes the subgraph to generate representations for both users and items. The user representation \( \mathbf{h}_u^{(k)} \) is iteratively refined through message passing over connected interactions.
    \item \textsc{Pair-specific interaction encoding}: The item representation \( \mathbf{h}_i^{(k)} \) is obtained directly from the same GNN model, ensuring that item embeddings are informed by user-specific interactions within the subgraph.
    \item \textsc{Score computation}: The final pair-wise recommendation score is determined by the dot product of the learned user and item representations:
    \begin{equation}
        y^{\text{(pair)}}(u, i) = \; <\mathbf{h}_u^{(k)}, \mathbf{h}_i^{(k)}>.
    \end{equation}
\end{enumerate}

\textbf{Two-tower representation score.}  
When the item \( i \) does not belong to the user’s local subgraph \( \mathcal{N}_k(u) \), a two-tower representation is employed. This approach allows for the recommendation of exploratory items, which may not be explicitly connected to the user but share latent similarities with items the user has interacted with. The two-tower representation is computed as follows:

\begin{enumerate}
    \item \textsc{Global user representation}: The user representation \( \mathbf{h}_u^{(k)} \) is still obtained from the GNN but without relying on the local subgraph structure of \( i \).
    \item \textsc{Static item embedding matrix}: Instead of deriving \( \mathbf{h}_i \) from the GNN, a pretrained item embedding matrix \( \mathbf{Q} \in \mathbb{R}^{|\mathcal{I}| \times d} \) is used, where each row corresponds to a static vector representation of an item.
    \item \textsc{Score computation}: The two-tower recommendation score is then computed as:
    \begin{equation}
        y^{\text{(tower)}}(v, w) = \; <\mathbf{h}_u^{(k)}, \mathbf{q}_i>.
    \end{equation}
\end{enumerate}

\textbf{Final fusion and adaptive weighting.}  
To dynamically balance familiar and exploratory recommendations, ContextGNN learns a fusion mechanism through a multi-layer perceptron (MLP) applied to the user representation. For each candidate item \( i \), the final recommendation score is computed as:

\begin{equation}
    y(u, i) =
    \begin{cases} 
    y^{\text{(pair)}}(u, i) + \text{MLP}_\theta(\mathbf{h}_u^{(k)}) & \text{if } i \in \mathcal{N}_k(u) \\
    y^{\text{(tower)}}(u, i) & \text{otherwise}.
    \end{cases}
\end{equation}

\section{Experimental study}
In this section, we present our experimental study to reproduce the results of ContextGNN on Gowalla, Yelp 2018, and Amazon Book against six state-of-the-art GNN-based recommender systems within \textsc{Elliot}. In the following, we describe the three datasets, the baselines we tested against ContextGNN, and the experimental settings to reproduce the results. Then, we report and discuss the obtained performance measures. 

\textbf{Datasets.} We use three recommendation datasets: Gowalla, Yelp 2018, and Amazon Book (\Cref{tab:datasets}). Gowalla~\cite{DBLP:conf/www/LiangCMB16} records the check-in history of users, where each check-in event corresponds to a location. Yelp 2018~\cite{DBLP:conf/sigir/0001DWLZ020} collects users' reviews about local business and derives from the 2018 Yelp challenge. Amazon Book~\cite{DBLP:conf/www/HeM16} is a product category from the popular Amazon dataset. 

\textbf{Baselines.} We select six state-of-the-art GNN-based recommendation systems following the same models selection in~\cite{DBLP:conf/recsys/AnelliMPBSN23}. Among them, NGCF~\cite{DBLP:conf/sigir/Wang0WFC19}, DGCF~\cite{DBLP:conf/sigir/WangJZ0XC20}, LightGCN~\cite{DBLP:conf/sigir/0001DWLZ020}, and SGL~\cite{DBLP:conf/sigir/WuWF0CLX21} represent pioneer recommendation systems built upon the message-passing graph neural networks schema. Conversely, UltraGCN~\cite{DBLP:conf/cikm/MaoZXLWH21} and GFCF~\cite{DBLP:conf/cikm/ShenWZSZLL21} settle as more advanced approaches surpassing the classical concept of message-passing to tackle known issues in the literature, such as oversmoothing. 

\textbf{Reproducibility details.} To rigorously reproduce the results of ContextGNN, we integrated its original code (accessible at this link\footnote{\url{https://github.com/kumo-ai/ContextGNN/tree/xinwei_add_static_data_and_model_v1}} for the model's version working on static link prediction) within \textsc{Elliot}. According to~\cite{DBLP:journals/corr/abs-2411-19513}, ContextGNN was only tested on Amazon Book, and no indication on the best hyper-parameter values was reported. Thus, to reproduce the results on Amazon Book and on the remaining two datasets (Gowalla and Yelp 2018) we carefully analyzed the original GitHub repository and emailed the authors to ask for more information. 

Based on their exhaustive answer, all model's hyper-parameters were set to the default values (the same on all datasets) as reported in this script\footnote{\href{https://github.com/kumo-ai/ContextGNN/blob/xinwei_add_static_data_and_model_v1/examples/static_example.py}{xinwei\_add\_static\_data\_and\_model\_v1/examples/static\_example.py}}. Conversely, for a specific parameter, \textsc{num\_neighbors}, we set it to different values on each dataset following this comment on GitHub\footnote{\url{https://github.com/kumo-ai/ContextGNN/pull/27\#issue-2527170237}}. Finally, for the three recommendation datasets, we selected the same dataset splitting as in~\cite{DBLP:conf/recsys/AnelliMPBSN23}, which corresponds to the one in the original baselines' papers.

\textbf{Results and discussion.} \Cref{tab:reproducibility} displays the results for our reproducibility analysis, where the values for all the baselines (apart from ContextGNN) were directly taken from~\cite{DBLP:conf/recsys/AnelliMPBSN23} as they were computed under the same experimental settings. 

On the one side, we confirm that the results for ContextGNN were closely reproduced on all datasets. Specifically, the reader might refer to the original paper~\cite{DBLP:journals/corr/abs-2411-19513} for the authors' results on Amazon Book while, for the other datasets, the reference is (again) at the GitHub comment reported above, under the sentence ``\textit{After removing the known labels from train set [...]}''. This was further confirmed by the authors during our email exchange.

On the other side, the outcomes demonstrate, in accordance with the original paper, that ContextGNN is capable of outperforming more traditional models based upon message-passing graph neural networks such as NGCF, DGCF, and (in some cases) LightGCN. However, the model still shows room for improvement when compared to more advanced and recent models, such as UltraGCN and GFCF. To this end, we maintain the importance of our proposed tool to run more extensive benchmarks for ContextGNN and, more generally, relational deep learning in the task of personalized item recommendation. 

\begin{table}[!t]
\centering
\caption{Statistics of the tested datasets.}
\label{tab:datasets}
\begin{tabular}{lrrrr}
\toprule
\textbf{Datasets} & \textbf{Users} & \textbf{Items} & \textbf{Interactions} & \textbf{Sparsity} \\
\cmidrule{1-5}
Gowalla & 29,858 & 40,981 & 1,027,370 & 0.9992 \\
Yelp 2018 & 31,668 & 38,048 & 1,561,406 & 0.9987 \\
Amazon Book & 52,643 & 91,599 & 2,984,108 & 0.9994 \\
\bottomrule
\end{tabular}
\end{table}

\begin{table}[t]
\centering
\caption{Experimental results for all selected GNN-based recommender systems on Gowalla, Yelp 2018, and Amazon Book. Apart from the results of ContextGNN, all other results are taken from the reproducibilty study in~\cite{DBLP:conf/recsys/AnelliMPBSN23}.}
\label{tab:reproducibility}
\begin{adjustbox}{max width=\columnwidth}
\begin{tabular}{lcccccc}
\toprule
\textbf{Models} & \multicolumn{2}{c}{\textbf{Gowalla}} & \multicolumn{2}{c}{\textbf{Yelp 2018}} & \multicolumn{2}{c}{\textbf{Amazon Book}} \\ \cmidrule(lr){2-3} \cmidrule(lr){4-5} \cmidrule(lr){6-7} 
& \multicolumn{1}{c}{Recall} & \multicolumn{1}{c}{nDCG} & \multicolumn{1}{c}{Recall} & \multicolumn{1}{c}{nDCG} & \multicolumn{1}{c}{Recall} & \multicolumn{1}{c}{nDCG} \\
\cmidrule{1-7} 
NGCF & 0.1556 & 0.1320 & 0.0556 & 0.0452 & 0.0319 & 0.0246 \\
DGCF & 0.1736 & 0.1477 & 0.0621 & 0.0505 & 0.0384 & 0.0295 \\
LightGCN & 0.1826 & \underline{0.1545} & 0.0629 & 0.0516 & 0.0419 & 0.0323 \\
SGL* & \multicolumn{1}{c}{---} & --- & 0.0669 & 0.0552 & 0.0474 & 0.0372 \\
UltraGCN & \textbf{0.1863} & \textbf{0.1580} & \underline{0.0672} & 0.0553 & \underline{0.0688} & 0.0561 \\
GFCF & \underline{0.1849} & 0.1518 & \textbf{0.0697} & \textbf{0.0571} & \textbf{0.0710} & \textbf{0.0584} \\
\cmidrule{1-7}
\rowcolor{blue!5!white} ContextGNN & 0.1712 & 0.1285 & 0.0543 & 0.0430 & 0.0455 & 0.0379 \\
\bottomrule
\multicolumn{7}{l}{\footnotesize\textit{*Results are not reported on Gowalla as SGL was not originally trained and tested on it~\cite{DBLP:conf/sigir/WuWF0CLX21}.}}
\end{tabular}
\end{adjustbox}
\end{table}

\section{How to benchmark ContextGNN within Elliot}

Running benchmarking analyses with ContextGNN in \textsc{Elliot} is quite straightforward and consists of four main steps that we outline next. Note that the running steps apply to the current version of the benchmarking tool. We acknowledge the existence of some limitations that we plan to address in future extensions of this work.

\noindent \textbf{1) Prepare your recommendation data.} To work properly, the framework needs (in the minimum setting) the following files: 

\begin{itemize}
    \item The list of users and items, two separate files formatted as tsv files with the original user (item) IDs and their numerical mapped IDs. The two files are named \textsc{user\_list.txt} and \textsc{item\_list.txt}, and their column headers are named \textsc{org\_id} and \textsc{remap\_id}.
    \item The training and test files (optionally, the validation file). They are formatted as tsv files where each row contains the user ID and the IDs of all the items the user has interacted with. Thus, the total number of rows will reflect the number of users in the dataset. All IDs in these files are intended to be the mapped IDs (i.e., reported in the \textsc{remap\_id} columns from the \textsc{user\_list.txt} and \textsc{item\_list.txt} files).
\end{itemize}  

\noindent \textbf{2) Prepare the dataset for \textsc{Elliot} and RelBench.} In this step, the input dataset is processed to be used by RelBench within \textsc{Elliot} by running the Python script \textsc{map\_rel\_bench.py}. 

\begin{tcolorbox}[colback=blue!5!white,colframe=black!75!black,title=\textsc{\bfseries Run the code!}]
\texttt{python map\_rel\_bench.py \textnormal{---}dataset <dataset\_name>}
\end{tcolorbox}

\noindent This script reads the files from the previous step and creates the following files:
\begin{itemize}
    \item Two tsv files \textsc{train\_elliot.tsv} and \textsc{test\_elliot.tsv}, formatted with two columns, one for the user and the other for the items, where each row corresponds to an interaction. These files are useful for \textsc{Elliot} to execute properly.
    \item Two tsv files, \textsc{src\_df.tsv} and \textsc{dst\_df.tsv}, that represent the user and item tables within the database, with a column for the user (item) IDs and the other containing a dummy timestamp value (as required in RelBench). Then, other two files are \textsc{train\_df.tsv} and \textsc{test\_df.tsv}, formatted with three columns, one indicating the user, the other the list of interacted items, and the final one (again) as a dummy timestamp value. Finally, a tsv file named \textsc{target\_table.tsv}, whose content is quite similar to that of \textsc{train\_elliot.tsv}, with an additional column for the dummy timestamp value. These five files are required by RelBench to execute properly.
\end{itemize}

\noindent \textbf{3) Prepare the configuration file for ContextGNN.} From this step on, the process to perform the benchmark experiments is exactly the same as in \textsc{Elliot}~\cite{DBLP:conf/recsys/AnelliMPBSN23}. While we invite the readers to refer to the original ContextGNN's paper~\cite{DBLP:journals/corr/abs-2411-19513} and code\footnote{\url{https://github.com/kumo-ai/ContextGNN}} for a proper explanation and setting of all the hyper-parameters, as well as the official \textsc{Elliot}'s documentation for further details regarding how to prepare a YAML configuration file\footnote{\url{https://elliot.readthedocs.io}}, here we report (as an example) the configuration file we used to reproduce the results of ContextGNN on Gowalla.

\begin{lstlisting}[language=yaml]
experiment:
  backend: pytorch
  data_config:
    strategy: fixed
    train_path: ../data/{0}/train_elliot.tsv
    test_path: ../data/{0}/test_elliot.tsv
  dataset: gowalla
  [...]
  models:
    external.ContextGNN:
      meta:
        hyper_opt_alg: grid
        verbose: True
        save_weights: False
        validation_rate: 20
        validation_metric: Recall@20
        restore: False
      lr: 0.001
      epochs: 20
      factors: 128
      batch_size: 128
      n_layers: 4
      aggr: sum
      channels: 128
      max_steps: 2000
      neigh: (16,16,16,16)
      seed: 42
\end{lstlisting}

\noindent \textbf{4) Run Elliot to benchmark ContextGNN.} If all the previous steps run smoothly, we are all set to execute the experiment by running:

\begin{tcolorbox}[colback=blue!5!white,colframe=black!75!black,title=\textsc{\bfseries Run the code!}]
\texttt{python start\_experiments.py \\\ \textnormal{---}dataset <dataset\_name> \textnormal{---}model <model\_name>}
\end{tcolorbox}

\noindent While the final results can be directly observed at the end of the output log produced by \textsc{Elliot}, a more convenient way is to search for the file \textsc{rec\_cutoff\_20\_relthreshold\_0\_<datetime>.tsv} at the path \textsc{results/<dataset\_name/performance/}.
\section{Conclusion and future work}
Relational deep learning (RDL) has settled as the de facto standard in deep learning for relational databases, leveraging the representational power of message-passing graph neural networks (GNNs) to extract meaningful data patterns from tables connected through primary-to-foreign key connections. Among its recent applications to real-world prediction tasks and datasets, this paradigm has shown promising performance trends also on more traditional static link prediction settings (aka personalized item recommendation) with the proposal of a novel recommendation model, namely, ContextGNN. Despite effective, we verified ContextGNN authors' claim, indicating potential room for improvements, especially in certain recommendation scenarios. To facilitate such a development, with this paper, we decided to integrate the original ContextGNN code within \textsc{Elliot}, a popular framework for rigorous and extensive reproducibility and benchmarking of recommender systems. On such a basis, we reproduced ContextGNN's performance on three recommendation datasets (Gowalla, Yelp 2018, and Amazon Book), against six state-of-the-art GNN-based recommender systems, confirming the trends observed by the authors in the original paper. Then, we showed how to easily run extensive benchmarking analyses for ContextGNN within the \textsc{Elliot} environment. With this paper and tool, we hope to foster future research on relational deep learning for personalized item recommendation. In this respect, we are thrilled to invite any contribution to improve, among others, the usability of the tool and its modularity to integrate the relational deep learning paradigm (especially through the RelBench package) more seamlessly within \textsc{Elliot} to facilitate more extensive benchmarking analyses. 

\bibliographystyle{ACM-Reference-Format}
\bibliography{main}

\end{document}